\documentclass[letterpaper, 10 pt, conference]{ieeeconf}

\IEEEoverridecommandlockouts
\usepackage{amsmath} 
\usepackage{amssymb}  
\usepackage{amsfonts}
\usepackage{xspace}
 \usepackage{epsfig}
\usepackage{graphicx}
\usepackage{epstopdf}
\usepackage{color}
\usepackage{subfigure}

\usepackage{color}

\usepackage{graphicx}
\usepackage{amsmath}

\usepackage{amsfonts}
\usepackage{mathrsfs}

\begin{document}

 \title{Quantum Master Equation and Filter for Systems Driven by Fields in a Single Photon State}

\author{John E.~Gough\thanks{J.~Gough is with the Institute for Mathematical and Physical Sciences,
Aberystwyth University, Ceredigion, SY23 3BZ, Wales. jug@aber.ac.uk Research supported by EPSRC.} 
\and
Matthew R.~James\thanks{M.~James is with the Centre for Quantum Computation and Communication Technology, School of
    Engineering, Australian
    National University, Canberra, ACT 0200,
    Australia. Matthew.James@anu.edu.au. Research supported by the
    Australian Research Council. Corresponding author.}
    \and
 Hendra I.~Nurdin\thanks{H.~Nurdin is with the School of
    Engineering, Australian
    National University, Canberra, ACT 0200,
    Australia. Hendra.Nurdin@anu.edu.au. Research supported by the
    Australian Research Council.}}

 \date{\today}

\maketitle
\begin{abstract}
The aim of this paper is  to determine quantum master and filter equations for systems coupled to continuous-mode single photon fields. The system and field are described using a quantum stochastic unitary model, where the continuous-mode single photon state for the field is determined by a wavepacket pulse shape. The master equation is derived from this model and is given in terms of a system of coupled equations.
The output field carries information about the system from the scattered photon, and is continuously monitored.
The quantum filter  is determined with the aid of an embedding of the system into a larger   system, and is given by a system of coupled stochastic differential equations. An example is provided to illustrate the main results.
\end{abstract}

\section{Introduction}
\label{sec:intro}

\begin{figure}[h]
\begin{center}
\includegraphics{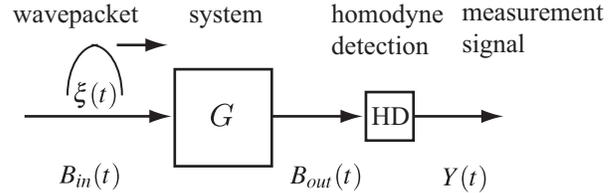} 
\end{center}
\caption{System $G$ coupled to a field in a continuous-mode single photon state $\vert
\protect\xi \rangle$ defined by the wavepacket $\protect\xi$. The output
field is continuously monitored by homodyne detection (assumed perfect) to produce a classical
measurement signal $Y(t)$.}
\label{fig:filter-one-1}
\end{figure}

In recent years single photon states of light have become increasingly
important due to applications in quantum technology, in particular, quantum
computing and quantum information systems, \cite{MHNWGBRH97}, \cite{GM08},
\cite{KLM01}, \cite{GRTZ02}, \cite{VWSRVSKW06}. For instance, the light may
interact with a system, say an atom, quantum dot, or cavity, and this system may be used
as a quantum memory, \cite{MHNWGBRH97}, or  to control
the pulse shape of the single photon state \cite{GM08}. Note that in practice one can consider different 
types of `single photon
states', including  the single photon state of a single mode of light  confined inside an optical cavity \cite{MHNWGBRH97}, or a single photon state superposed over a continuum of modes of a travelling
field (i.e., colloquially, a ``flying'' single photon state) referred to as a continuous-mode or multimode single photon state \cite{RL00}, \cite{GM07}, \cite{GM08}. In this
paper we will be interested exclusively with the latter kind of single photon state and therefore from this point on when we say
`single photon' state we specifically mean the continuous-mode single photon state of a travelling field.  When light interacts
with a quantum system, information about the  system
is contained in the scattered light. This information may be useful for
monitoring the behavior of the  system, or for controlling it. The
topic of this paper concerns the extraction of information from the
scattered light when the incoming light is placed in a single photon state,
denoted $\vert \xi \rangle$, as illustrated in Figure \ref{fig:filter-one-1}.

The problem of extracting information from continuous measurement of the
scattered light is a problem of \emph{quantum filtering}, \cite{BB91}, \cite
{VPB92}, \cite{VPB92a}, \cite{HC93}, \cite{WM93}, \cite{AB03},  \cite{BHJ07}, \cite{WM10}.
The current state of the art for quantum filtering considers incoming light
in a vacuum or other Gaussian state, with quadrature or counting measurements.
Single photon states of light are highly non-classical, and are
fundamentally different from Gaussian  states. At present, to the best of our knowledge,  there are no
filtering results for systems driven by single photon fields. In view of the
increasing importance of single photon states of light, the purpose of this
paper is to solve a quantum filtering problem for systems driven by single
photon fields. As a by-product, we obtain a master equation for the system.
A significant feature of the master equation and the quantum filter is that
they are both given by a system of coupled equations, not a single equation
as in the vacuum case.  This reflects the non-Markovian character
of systems driven by single photon fields.

The paper is organized as follows. The filtering problem to be solved is formulated in Section \ref{sec:problem}.  The master equation is
derived in Section \ref{sec:master} using the model presented in Section \ref
{sec:problem}. This leads naturally to Section \ref{sec:photon-signal}, where
the system is embedded in a larger  Markovian model, using a signal generator model.  The single photon filter is presented in Section \ref{sec:photon-filter}, and an example is discussed in Section \ref{sec:photon-eg}.

\emph{Notation:} We use the standard Dirac notation $\vert \psi \rangle$ to
denote state vectors (vectors in a Hilbert space) \cite{EM98}, \cite{AFP09}.
The superscript $^\ast$ indicates Hilbert space adjoint or complex
conjugate. The inner product of state vectors $\vert \psi_1\rangle$ and $\vert \psi_2 \rangle$  is denoted $\langle \psi_1 \vert
\psi_2 \rangle$. The expected value of an operator $X$ when the system is in state $\vert \psi \rangle$ is denoted $\mathbb{E}_\psi[ X ] = \langle \psi \vert X \vert \psi \rangle$. For operators $A$ and $B$ we write
$
\langle A, B \rangle = \mathrm{tr}[ A^\ast B ].
$

\section{Problem Formulation}
\label{sec:problem}

We consider a quantum system $G$ coupled to a quantum field $B_{in}$, as shown in
Figure \ref{fig:filter-one-1}. The interaction of $B_{in}$ with $G$ produces the output field $B_{out}$. The input field is placed in
a single photon state, denoted using Dirac's notation as $\vert \xi \rangle$, where $\xi$ is a complex valued function such that $\int_0^\infty \vert
\xi(s) \vert^2 ds =1$. As illustrated in Figure \ref{fig:filter-one-1}, the
wavepacket $\vert \xi \rangle$ interacts with the quantum system $G$, and
the results of this interaction provide information about the system that
may be obtained through continuous measurement of an observable $Y(t)$ of
the output field $B_{out}(t)$. The filtering problem of interest in this
paper is to determine the conditional state from which estimates $\hat X(t)$
of system operators $X$ may be determined at time $t$ based on knowledge of
the observables $\{ Y(s)$, $0 \leq s \leq t \}$.

In what follows the system $G$ is assumed to be defined on a Hilbert space $%
\mathfrak{H}_S$, with an initial state denoted $\vert  \eta \rangle   \in %
\mathfrak{H}_S$. The input field $B_{in}$ is described in terms of
annihilation $B(\xi)$ and creation $B^\ast(\xi)$ operators defined on a symmetric (Boson) Fock
space $\mathfrak{F}$, \cite[Chapter II]{KRP92}, \cite[Section 4]{BHJ07}.
The continuous-mode single photon state is defined on the symmetric Fock space by \cite[sec. 6.3]{RL00}, \cite[Section 14.2]{GM07}, \cite[eq. (9)]{GM08}
\begin{equation}
\vert \xi \rangle = B^\ast(\xi) \vert \phi \rangle,  
\label{eq:xi-create}
\end{equation}
where $\vert \phi \rangle$ is the vacuum state of the field. Expression (\ref{eq:xi-create}) says that  the single photon wavepacket is created from the vacuum using the field operator $B^\ast(\xi)$.  

The Hilbert space for the composite system is
\begin{equation*}
\mathfrak{H} = \mathfrak{H}_S \otimes \mathfrak{F}= \mathfrak{H}_S \otimes %
\mathfrak{F}_{t]} \otimes \mathfrak{F}_{(t},
\end{equation*}
where here we have exhibited the continuous temporal tensor product
decomposition of the Fock space $\mathfrak{F}=\mathfrak{F}_{t]} \otimes %
\mathfrak{F}_{(t}$ into past and future components, which is of basic
importance in what follows.  We use the notation $\mathbb{E}$ to denote
quantum expectation, usually with a subscript to denote the state being
used. In particular, we write
\begin{equation}
\mathbb{E}_{11} [ X \otimes F ] = \langle \eta \xi \vert  (X \otimes F) \vert \eta
\xi \rangle = \langle \eta  \vert X \vert \eta \rangle \langle \xi \vert  F  \vert \xi \rangle
\end{equation}
for the expectation with respect to the product state $\vert \eta \xi \rangle$, where the field is in the single
photon state.  Here and in what follows  $X$ is a bounded system operator acting on $\mathfrak{H}_S$,
and $F$ is a field operator acting on the Fock space $\mathfrak{F}$.
Similarly, we may define the expectation when the field is in the vacuum
state,
\begin{equation}
\mathbb{E}_{00} [ X \otimes F ] = \langle  \eta \phi \vert  (X \otimes F) \vert \eta
\phi  \rangle = \langle \eta \vert X \vert \eta \rangle \langle \phi \vert  F \vert \phi  \rangle .
\end{equation}
We will also have need for the cross-expectations
\begin{eqnarray}
\mathbb{E}_{10}[ X \otimes F ] = \langle \eta\xi \vert  (X \otimes F)  \vert \eta\phi
\rangle, 
\notag \\
  \mathbb{E}_{01}[ X \otimes F ] = \langle \eta\phi \vert
(X \otimes F) \vert  \eta \xi \rangle.
\end{eqnarray}

A crucial difference between the single photon state and the vacuum state is
that the later state factorizes $\vert \phi \rangle =\vert \phi_{t]} \rangle
\otimes \vert \phi_{(t} \rangle $ with respect to the temporal factorization
$\mathfrak{F}=\mathfrak{F}_{t]} \otimes \mathfrak{F}_{(t}$ of the Fock
space, with $\vert \phi_{t]} \rangle \in \mathfrak{F}_{t]} $ and $\vert
\phi_{(t} \rangle \in \mathfrak{F}_{(t}$, while the former does not. Rather,
we have
\begin{equation}
\vert \xi \rangle = B^\ast(\xi) \vert \phi \rangle = \vert \xi_{t]} \rangle
\otimes \vert \phi_{(t} \rangle + \vert \phi_{t]} \rangle \otimes \vert
\xi_{(t} \rangle ,  \label{eq:factor-additive-1}
\end{equation}
where
\begin{equation}
\vert \xi_{t]} \rangle = B^{-\ast}_t(\xi) \vert \phi_{t]} \rangle, \ \text{%
and} \ \vert \xi_{(t} \rangle = B^{+\ast}_t(\xi) \vert \phi_{(t} \rangle ,
\end{equation}
and $B^-_t(\xi) = B(\xi \chi_{[0,t]})$, $B^+_t(\xi) = B(\xi \chi_{(t,\infty)})$, $B(\xi) = B^-_t(\xi) + B^+_t(\xi)$.
 Here, $\chi_{[0,t]}$ is the indicator function for the time interval $[0,t]$%
. Note that while $\vert \xi \rangle$ has unit norm, we have
\begin{equation}
\parallel \vert \xi_{t]} \rangle \parallel^2 = \int_0^t \vert \xi(s) \vert^2
ds, \ \text{and} \ \parallel \vert \xi_{(t} \rangle \parallel^2 =
\int_t^\infty \vert \xi(s) \vert^2 ds .
\end{equation}

A consequence of the additive decomposition (\ref{eq:factor-additive-1})   is the following. Let $K(t)$ be a bounded operator acting on the full Hilbert space $\mathfrak{H}$ that is adapted, i.e. $K(t)$ acts trivially on $\mathfrak{F}_{(t}$, the field in the future.
Then the expectation with respect to the single photon field may be
expressed in terms of the vacuum state as follows:
\begin{eqnarray}
\mathbb{E}_{11} [ K(t) ] &=& \mathbb{E}_{00} [ B^-_t(\xi) K(t)
B^{-\ast}_t(\xi) + w(t) K(t) ]  
\label{eq:xi-phi}
\end{eqnarray}
where $w(t) = \int_t^\infty \vert \xi(s) \vert^2 ds$.

The dynamics of the system will be described using the quantum stochastic
calculus, \cite{HP84}, \cite{GC85}, \cite{KRP92}, \cite{GZ00}, \cite{BHJ07}.
 Quantum stochastic integrals are
defined in terms of fundamental field operators $B(t)$, $B^\ast(t)$ and $%
\Lambda(t)$, \cite[Chapter II]{KRP92}, \cite[Section 4]{BHJ07}.\footnote{%
In terms of annihilation and creation white noise operators $b(t), b^\ast(t)$
that satisfy singular commutation relations $[b(s), b^\ast(t)]=\delta(t-s)$,
the fundamental field operators are given by $B(t) = \int_0^t b(s) ds$, $%
B^\ast(t)= \int_0^t b^\ast(s) ds$, and $\Lambda(t)= \int_0^t b^\ast(s) b(s)
ds$. Also, we may write $B(\xi) = \int_0^\infty \xi^\ast(s) dB(s)$.} The
non-zero Ito products for the field operators are
\begin{eqnarray}
dB(t) dB^\ast(t) = dt, \ \ dB(t) d\Lambda(t) = dB(t), \
\notag \\
 d\Lambda(t)d\Lambda(t) = d\Lambda(t), \ \ d\Lambda(t) dB^\ast(t)=dB^\ast(t). \label{eq:Ito-sp}
\end{eqnarray}

The dynamics of the composite system is described by a unitary $U(t)$
solving the Schr\"{o}dinger equation, or quantum stochastic differential equation (QSDE),
\begin{eqnarray}
dU(t) = \{ (S-I)d\Lambda(t) + L dB^\ast (t)- L^\ast SdB(t) 
\notag \\
- (\frac{1}{2}
L^\ast L +iH )dt \} U(t),  \label{eq:unitary}
\end{eqnarray}
with initial condition $U(0)=I$. Here, $H$ is a fixed self-adjoint operator
representing the free Hamiltonian of the system, and $L$ and $S$ are system
operators determining the coupling of the system to the field, with $S$
unitary. In this paper, for simplicity we assume that the parameters $G=(S,L,H)$ are bounded operators
on the system Hilbert space $\mathfrak{H}_S$.  

A system operator $X$ at time $t$ is given in the Heisenberg picture by $%
X(t)=j_{t}( X) =U( t) ^{\ast } ( X\otimes I ) U( t) $ and it follows from the
quantum Ito calculus that
\begin{eqnarray}
dj_{t}( X) &=&j_{t}(S^{\ast }XS-X) d\Lambda ( t) +j_{t}(S^{\ast }[X,L]) dB(t) ^{\ast }  
\notag \\
&& 
+j_{t}([L^{\ast },X]S) dB( t) +j_{t}(\mathcal{G}(X)) dt ,
\label{eq:X-dyn}
\end{eqnarray}
where
\begin{equation}
\mathcal{G}(X)= \mathcal{L}_L(X)-i[ X,H] ,
\end{equation}
and
\begin{equation}
\mathcal{L}_L(X)=\frac{1}{2}L^{\ast }[X,L]+\frac{1}{2}[L^{\ast },X]L.
\end{equation}
and
The map $X\mapsto \mathcal{G} (X)$ is known as the \emph{Lindblad generator},
while the quartet of maps $X \mapsto \mathcal{G} (X), S^\ast XS -X, \,
S^\ast [X,L ], \, [L^\ast , X] S$ are known as \emph{Evans-Hudson maps}.

The output field is defined by $B_{out}(t) = U^\ast(t) B(t) U(t)$.\footnote{%
Recall $B(t)=B_{in}(t)$ is the input field.} In this paper we consider the
output field observable $Y(t)$ defined by
\begin{equation}
Y(t) = U^\ast(t) Z(t) U(t) , 
 \label{eq:Y-out}
\end{equation}
where
\begin{equation}
Z(t)= B(t)+ B^\ast(t), 
 \label{eq:Z-def}
\end{equation}
is a quadrature  observable of the input field.  Note that both $Z(t)$ and $Y(t)$ are
self-adjoint and self-commutative: $[Z(t), Z(s)]=0$ and $[Y(t), Y(s)]=0$. We
write $\mathscr{Z}_t$ and $\mathscr{Y}_t$ for the subspaces of commuting
operators generated by the observables $Z(s)$, $Y(s)$, $0\leq s \leq t$,
respectively.\footnote{$\mathscr{Z}_t$ and $\mathscr{Y}_t$ are commutative
von Neumann algebras. They are also filtrations, e.g. $\mathscr{Z}_{t_1}
\subset \mathscr{Z}_{t_2}$ whenever $t_1 <t_2$.} They are related by the
unitary rotation $\mathscr{Y}_t = U^\ast(t) \mathscr{Z}_t U(t)$. Physically,
$Y(t)$ may represent the integrated photocurrent arising in an idealized (perfect)  homodyne
photodetection scheme, as in Figure \ref{fig:filter-one-1}. For further information on homodyne detection, we refer the reader to the literature;  for example, \cite{BR04},  \cite{AB03}, 
\cite{WM10}, \cite{GG01}. In particular, \cite{GG01} considers pulsed homodyne detection for fields in a continuous-mode $n$-photon state, which
includes the single photon state as a special case.

The primary goal of this paper is to determine the \emph{quantum filter} for
the quantum conditional expectation (see, e.g. \cite[Definition 3.13]{BHJ07}%
)
\begin{equation}
\hat X(t) = \mathbb{E}_{11}[ X(t) \, \vert \, \mathscr{Y}_t ] .
\label{eq:cond-exp}
\end{equation}
This conditional expectation is well defined, since $X(t)$ commutes with the
subspace $\mathscr{Y}_t $ (non-demolition condition). The conditional
estimate $\hat X(t)$ is affiliated to $\mathscr{Y}_t$ (written in abbreviated fashion as  $\hat X(t) \in \mathscr{Y}_t$)
and is characterized by the requirement that
\begin{equation}
\mathbb{E}_{11} [ \hat X(t) K ] = \mathbb{E}_{11} [ X(t) K ]
\label{eq:c-exp-def}
\end{equation}
for all $K \in \mathscr{Y}_t$.

\section{Master Equation}
\label{sec:master}

Before deriving the quantum filter, we work out dynamical equations for the
unconditioned single photon expectation. Such equations are often called
\emph{master equations} and are of fundamental importance, and arise in
\emph{Markovian} models of open quantum systems, \cite{GZ00}, \cite{KRP92},
\cite{HC93}, \cite{WM10}. Master equations are analogous to the Fokker-Plank
equations for classical diffusion processes.  Note that the master equations 
for systems driven by a single photon field have previously been derived by other means in
\cite{GEPZ98}, although we only became aware of this after this work was completed.

When the field is in the vacuum state $\vert \phi \rangle$, the joint
system-field state evolves according to $U(t) \vert \eta \phi \rangle$, and
the system density operator $\rho^{00}(t)$ is defined by $\langle
\rho^{00}(t) , X \rangle = \langle \eta\phi, U^\ast(t)( X \otimes I) U(t)
\eta\phi \rangle = \mathbb{E}_{00}[ X(t) ]$. It is well-known \cite{KRP92},
\cite{HP84}, \cite{GZ00} that $\rho^{00}(t)$ satisfies the master equation
\begin{equation}
\dot \rho^{00}(t ) = \mathcal{G}^\ast( \rho^{00}(t) ), 
 \label{eq:master-vac}
\end{equation}
where
\begin{equation}
\mathcal{G}^\ast(\rho) = L \rho L^\ast - \frac{1}{2} \rho L^\ast L - \frac{1%
}{2} L^\ast L \rho +i [\rho, H].  \label{eq:Lstar-rho}
\end{equation}
The master equation (\ref{eq:master-vac}) is readily determined from the
Heisenberg evolution (\ref{eq:X-dyn}) by taking expectations with respect to
the vacuum state and appropriately collecting terms. Note that the unitary operator $S$ appearing in the
Schr\"{o}dinger equation (\ref{eq:unitary}) does not appear in the master
equation (\ref{eq:master-vac}).

Now suppose that the field is in a single photon state $\vert \xi \rangle$,
in which case the density operator $\rho^{11}(t)$ is defined by $\langle
\rho^{11}(t) , X \rangle = \langle \eta\xi, U^\ast(t)( X \otimes I) U(t)
\eta\xi \rangle = \mathbb{E}_{11}[ X(t) ]$, which involves expectation with
respect to the single photon field. 
Using equation  (\ref{eq:X-dyn})  and  the relations
\begin{equation}
 dB(t) \vert \xi \rangle  = \xi(t) \vert  \phi  \rangle, \ \ d\Lambda(t) \vert \xi \rangle = \xi(t) dB^\ast(t) \vert  \phi  \rangle,
\label{eq:xi-annihilate-dB}
\end{equation}
we calculate that
\begin{eqnarray}
&&  \frac{d}{dt} \mathbb{E}_{\eta\xi} [ X(t) ] =  \frac{d}{dt}  \mathbb{E}_{11}[  X(t)]
\label{eq:master-photon-1} \\
&    & =
\mathbb{E}_{11}[  \mathcal{G} (X(t)) ]
\notag \\
&& +
\mathbb{E}_{01}[  S^\ast(t)[X(t),L(t)]   ] \xi^\ast(t)
\notag \\
&&
+ \mathbb{E}_{10}[  [L^\ast(t), X(t)] S(t)    \xi(t)
\notag \\
&& + \mathbb{E}_{00}[  (S^\ast(t) X(t) S(t) - X(t))   ]
\vert \xi(t) \vert^2. 
\notag
\end{eqnarray}
Notice that the right hand side of (\ref{eq:master-photon-1}) includes a vacuum expectation, as well as cross terms involving single photon and vacuum states. The system driven by the single photon field is not Markovian, in contrast to the vacuum case.

In view of this, we define
\begin{eqnarray}
\mu^{11}_t(X) = \mathbb{E}_{11}[  X(t) ], \ \ \mu^{10}_t(X) = \mathbb{E}_{10}[   X(t) ], 
\notag \\
\mu^{ 0 1}_t(X) = \mathbb{E}_{01}[  X(t) ] \ \ \mu^{00}_t(X) = \mathbb{E}_{00}[   X(t) ]. \  \
\label{eq:mu-photon-def}
\end{eqnarray}
Consequently, the master equation in Heisenberg form for the system  
when the field is in the single photon state $\vert \xi \rangle$ is given by
the system of equations
\begin{eqnarray}
\dot{\mu}^{11}_t (X) &=& \mu^{11}_t(\mathcal{G}(X)) + \mu^{01}_t( S^\ast
[X,L] ) \xi^\ast(t) 
\label{eq:rho-dyn-a-11}  \\
&& 
+ \mu^{10}_t( [L^\ast, X] S ) \xi(t)  
\notag \\
&&
+ \mu^{00}_t( S^\ast X S - X) \vert \xi(t) \vert^2,
\notag \\
\dot{\mu}^{10}_t (X) &=& \mu^{10}_t(\mathcal{G}(X)) + \mu^{00}_t( S^\ast [X,
L] ) \xi^\ast(t) ,  \label{eq:rho-dyn-a-10} \\
\dot{\mu}^{01}_t (X) &=& \mu^{01}_t(\mathcal{G}(X)) + \mu^{00}_t( [L^\ast,
X] S ) \xi(t) ,  \label{eq:rho-dyn-a-01} \\
\dot{\mu}^{00}_t (X) &=& \mu^{00}_t(\mathcal{G}(X)) .
\label{eq:rho-dyn-a-00}
\end{eqnarray}
The initial conditions are
\begin{equation}
\mu^{11}_0(X)= \mu^{00}_0(X)= \langle \eta, X \eta \rangle, \ \
\mu^{10}_0(X)= \mu^{01}_0(X)=0.
\end{equation}

In order to obtain a Schr\"{o}dinger form of the master equations, we define
(generalized) density operators $\rho^{jk}(t)$ by
\begin{equation}
\langle \rho^{jk}(t), X \rangle = \mu^{jk}_t(X) .
\end{equation}
The operators $\rho^{jk}(t)$ enjoy the symmetry properties
\begin{equation}
\rho^{00\ast}(t) = \rho^{00}(t), \ \ \rho^{01\ast}(t) = \rho^{10}(t), \ \
\rho^{11\ast}(t) = \rho^{11}(t) .
\end{equation}
 The master equation in Schr\"{o}dinger form for the
system  when the field is in the single photon state $\vert \xi \rangle$
is given by the system of equations
\begin{eqnarray}
\dot \rho^{11} (t) &=& \mathcal{G}^\ast( \rho^{11} (t) ) + [ S\rho^{01}(t),
L^\ast] \xi(t) 
\notag \\
&& + [ L , \rho^{10}(t)  S^\ast  ] \xi^\ast(t)  \notag \\
&& + (S \rho^{00}(t) S^\ast -\rho^{00}(t) ) \vert \xi(t) \vert^2,  \label{eq:rho-dyn-11}
\\
\dot \rho^{10} (t) &=& \mathcal{G}^\ast( \rho^{10} (t) ) + [ S \rho^{00}(t),
L^\ast ] \xi(t) ,  \label{eq:rho-dyn-10} \\
\dot \rho^{01}(t) &=& \mathcal{G}^\ast( \rho^{01}(t) ) + [ L,
\rho^{00} (t) S^\ast  ] \xi^\ast(t) ,  \label{eq:rho-dyn-01} \\
\dot \rho^{00} (t) &=& \mathcal{G}^\ast( \rho^{00}(t) ) .
\label{eq:rho-dyn-00}
\end{eqnarray}
The initial conditions are
\begin{eqnarray}
\rho^{11}(0) = \rho^{00}(0) = \vert \eta \rangle \langle \eta \vert, \ \
\rho^{10}(0) = \rho^{01}(0) = 0 .  
\label{eq:rho-jk-init}
\end{eqnarray}
 An example of the master equation is presented in Section \ref{sec:photon-eg}.

\section{Single Photon Signal Model}
\label{sec:photon-signal}

In Section \ref{sec:master} we saw that the master equation for the  system $G$
driven by a single photon field is non-Markovian, and the equations derived
suggest the possibility of embedding the system and field in a larger
 system $\tilde G$.
  Indeed, Markovian embeddings were used in \cite{HPB04} to
derive quantum trajectory equations for a class of non-Markovian master
equations. In engineering and statistics, it is common practice to use \lq{generating filters}\rq \ driven by white noise to represent colored noise.

\begin{figure}[h]
\begin{center}
\includegraphics[scale=0.9]{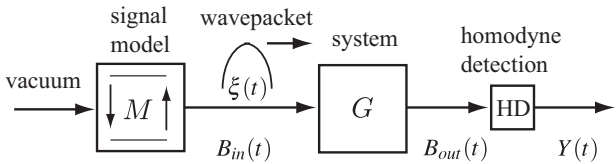}
\caption{An ancilla system $M$ (a two-level system) is used as a signal model or \lq{generating  filter}\rq. The ancilla system $M$ is driven by vacuum (quantum white noise), and produces a single-photon output. The cascade circuit illustrated in this figure is equivalent to the circuit of Fig. \ref{fig:filter-one-1}.}
\label{fig:system-signal}
\end{center}
\end{figure}

In this section we construct a generating  filter $M=(I,L_0, H_0)$ (an open quantum system on a Hilbert space $\mathfrak{H}_0$) driven by vacuum  to represent the single photon field, Fig. \ref{fig:system-signal}. Here, the ancilla parameters $M=(I,L_0, H_0)$  are to be determined.
This results in an extended system $\tilde G$ defined on the Hilbert space $\mathfrak{H}_0 \otimes \mathfrak{H}$ (cascade, or series connection, \cite{GJ09}) driven by vacuum, with parameters given by
\begin{equation}
\tilde G= G  \triangleleft M = (S, L+SL_0, H +H_0+ \mathrm{Im}[L^\ast  S  L_0]) ,
\label{series-product}
\end{equation}
 from which  the master equation and quantum filter equations (Section \ref{sec:photon-filter}) can be obtained.  

Specifically, 
we seek an ancilla system $M$  
initialized in a state $\vert \gamma \rangle$ and driven by vacuum that provides a unitary dilation of the single photon driven system. This means that 
given our system $G=(S,L,H)$, 
can we find an  ancilla system $M$ with state $\vert \gamma \rangle$ and an ancilla operator $R$ such that if $\tilde U(t)$ is the unitary for $\tilde G= G \triangleleft M$ then
\begin{equation}
\mathbb{E}_{\eta\xi}[ X(t) ] =  \mathbb{E}_{ \gamma\eta\phi} [ \tilde U^\ast(t) (R \otimes X ) \tilde U(t) ] .
\label{eq:photon-represent-1}
\end{equation}
Here, $X(t) = U^\ast(t) X U(t)$, where $U(t)$ is the unitary for $G$. Equation (\ref{eq:photon-represent-1}) means that the effect of the single photon field on the system is equivalent to the effect of the cascade of the  ancilla on the system. 
The left hand side of (\ref{eq:photon-represent-1}) is a quantum expectation with respect to the state $\vert \eta \xi \rangle = \vert \eta \rangle \otimes \vert \xi \rangle$ with the field in the single photon state $\vert \xi \rangle$, while the right hand side involves
 quantum expectation in the extended system with respect to the initial state $\vert \gamma \eta    \phi \rangle = \vert \gamma \rangle \otimes \vert \eta \rangle \otimes \vert \phi \rangle$, with the field in the vacuum state $\vert \phi \rangle$.

In order to fulfil the requirement (\ref{eq:photon-represent-1}), we consider the time derivative of both sides of (\ref{eq:photon-represent-1}) and determine the unknown signal model parameters by comparison.  The time derivative of the left hand side of (\ref{eq:photon-represent-1}) 
is given by (\ref{eq:master-photon-1}) or equations (\ref{eq:rho-dyn-a-11})-(\ref{eq:rho-dyn-a-00})  above, while the right hand derivative can be computed from the Lindblad generator   $\mathcal{G}_{G \triangleleft M}(A \otimes X)$ for the extended system  
$\tilde G = G \triangleleft M$,
\begin{eqnarray}
&& \frac{d}{dt} \mathbb{E}_{ \gamma\eta \phi} [ \tilde U^\ast(t) ( A \otimes X) \tilde U(t) ] 
\notag
\\
&=& \mathbb{E}_{ \gamma\eta \phi}[ \tilde U^\ast(t) \mathcal{G}_{G \triangleleft M}(A \otimes X)\tilde U(t) ],
\label{eq:photon-rep}
\end{eqnarray}
where
\begin{eqnarray}
  \mathcal{G}_{G \triangleleft M}(A \otimes X) &=&  A \otimes \mathcal{G}_G(X) + \mathcal{L}_{L_0}(A) \otimes X 
\notag \\
&& + 
L^\ast_0 A \otimes S^\ast [ X, L] + A L_0 \otimes [L^\ast, X]  S
\notag \\
&& 
+ L_0^\ast L_0 \otimes (S^\ast X S - X)
\end{eqnarray}
for any ancilla operator $A$ and system operator $X$.

Considering
the definitions (\ref{eq:mu-photon-def}) of $\mu^{jk}_t(X)$ above, we must find ancilla operators $Q_{jk}$ and (non-vanishing) weighting functions $w_{jk}(t)$ such that
\begin{equation}
\mu^{jk}_t(X) = \frac{ \tilde\mu_t( Q_{jk} \otimes X) }{ w_{jk}(t) },
\label{eq:claim-mujk}
\end{equation}
where
\begin{equation}
\tilde \mu_t( Q_{jk} \otimes X) = \mathbb{E}_{ \gamma\eta\phi} [ \tilde U^\ast(t) (Q_{jk} \otimes X ) \tilde U(t) ] .
\end{equation}
Now
\begin{eqnarray}
\frac{d}{dt} \frac{ \tilde\mu_t( Q_{jk} \otimes X) }{ w_{jk}(t) } &=& 
\frac{ \tilde \mu_t( \mathcal{G}_{G \triangleleft M}(Q_{jk} \otimes X))}{ w_{jk}(t) }
\notag \\
&&- \frac{ \tilde \mu_t( Q_{jk} \otimes X) }{ w_{jk}(t) } 
\frac{ \dot w_{jk}(t) }{w_{jk}(t) } .
\end{eqnarray}
Comparing this expression with equations (\ref{eq:rho-dyn-a-11})-(\ref{eq:rho-dyn-a-00}) we find that (\ref{eq:claim-mujk}) and (\ref{eq:photon-represent-1}) are 
 satisfied if we choose the ancilla to be a two-level system with state $\gamma = \vert e_1 \rangle$ (excited state), $R=Q_{11}=I$, $Q_{10}= \sigma_-$, $Q_{01}=\sigma_+$, $Q_{00}=n$, $w_{11}(t)=1$, $w_{10}(t)=w_{01}(t) = \sqrt{w(t)}$, $w_{00}(t) = w(t)=\int_t^\infty \vert \xi(s) \vert^2 ds$, $L_0 = \xi(t) \sigma_-/\sqrt{w(t)}$.

The signal model ancilla system is therefore 
\begin{equation}
M=(I, \frac{\xi(t) \sigma_-}{\sqrt{w(t)}},0) ,
\label{eq:photon-signal-model}
\end{equation}
 and so the extended system is  
 \begin{eqnarray}
&\tilde G= G \triangleleft M 
 \label{eq:extend-photon} \\
&= (S, L+ \frac{\xi(t)}{\sqrt{w(t)}} S\sigma_- , H+\frac{\xi(t)}{\sqrt{w(t)}}\mathrm{Im}(L^\ast S \sigma_-)) .
\notag
\end{eqnarray}

{\em Remark.}  The output state of the generating filter $M$ can be understood as follows. If $V(t)$ denotes the unitary for $M$ driven  by vacuum, and if the  initial state is $\vert \psi(0)  \rangle = \vert e_1 \rangle \otimes \vert \phi \rangle$, then the state $\vert \psi(t) \rangle = V(t)  \vert \psi(0)  \rangle$ satisfies the Schrodinger equation 
\begin{equation}
 d| \psi(t) \rangle  =\left[ \frac{\xi(t)}{\sqrt{w(t)}}
 \sigma_{-}dB^{\ast}(t) -
 \frac{1}{2} \frac{\vert \xi(t) \vert^2}{w(t)}
 \sigma _{+}\sigma _{-}dt\right] \,| \psi(t) \rangle 
\end{equation}
(since $dB(t) \vert \phi \rangle =0$). It
 is an elementary calculation to see that this has the exact solution
\begin{equation}
| \psi(t) \rangle  =\sqrt{w(t)}
| e_1
\otimes \phi \rangle +| e_0  \otimes B_{t}^{-\ast
}(\xi )\phi \rangle   ;
\label{two_level_entangled}
\end{equation}
cf. (\ref{eq:xi-phi}).
 Since $\xi$ is square integrable, $\vert \psi(t) \rangle$ approaches the state  $\vert e_0 \otimes  \xi \rangle$ as $t \to \infty$. Therefore the state of the field produced by the generating filter $M$ approaches the single photon state $\vert \xi \rangle$ asymptotically.

\section{Single Photon Filter}
\label{sec:photon-filter}

The quantum filter $\pi_t(X) =  \mathbb{E}_{\eta\xi}[ X(t) \vert Y(s), 0 \leq s \leq t ]$ 
for the system $G$ driven by a single photon field may now be obtained from the quantum filter $\tilde \pi_t(A \otimes X) = \mathbb{E}_{\gamma\eta \phi}[ \tilde U^\ast(t) (A \otimes X) \tilde U(t) \vert I\otimes Y(s), 0 \leq s \leq t]$  for the extended system $\tilde G=G \triangleleft M$ driven by vacuum \cite{BHJ07}, with the extended system parameters given by 
(\ref{eq:extend-photon}) (see the Appendix). Using the definition of conditional expectation (see \cite{BHJ07} and the Appendix),  it follows that 
\begin{eqnarray}
I\otimes \pi_t(X)  =
\tilde \pi_t( I \otimes X) .
\end{eqnarray}
If  we define
\begin{equation}
I \otimes \pi^{jk}_t(X) = \frac{ \tilde\pi_t( Q_{jk} \otimes X) }{ w_{jk}(t) },
\label{eq:pi-jk-photon}
\end{equation}
we obtain the system of equations
\begin{eqnarray}
&&  d\pi^{11}_t (X) 
  \notag
  \\ &=&  \bigl(\pi^{11}_t(\mathcal{G}(X)) + \pi^{01}_t( S^\ast [X,L] )
\xi^\ast(t)
\notag \\
&& 
 + \pi^{10}_t( [L^\ast, X] S ) \xi(t)  
+ \pi^{00}_t( S^\ast X S - X) \vert \xi(t) \vert^2\bigr)dt  \notag \\
&& + ( \pi^{11}_t( XL + L^\ast X) + \pi^{01}_t(S^\ast X) \xi^\ast(t) +
\pi^{10}_t( XS) \xi(t)  \notag \\
&& - \pi^{11}_t(X) \mathcal{K}_t(X)
)dW(t),  
\label{eq:pi-dyn-a-11} 
\end{eqnarray}
\begin{eqnarray}
d\pi^{10}_t (X) &=& \bigl( \pi^{10}_t(\mathcal{G}(X)) + \pi^{00}_t( S^\ast [X, L]
) \xi^\ast(t) \bigr)dt  \notag \\
&& +  ( \pi^{10}_t( XL + L^\ast X) + \pi^{00}_t(S^\ast X) \xi^\ast(t)  \notag
\\
&& - \pi^{10}_t(X)  \mathcal{K}_t(X)
  )dW(t),  
\label{eq:pi-dyn-a-10} 
\end{eqnarray}
\begin{eqnarray}
d\pi^{00}_t (X) &=& \pi^{00}_t(\mathcal{G}(X)) dt + ( \pi^{00}_t( XL +
L^\ast X)  \notag \\
&& - \pi^{00}_t(X)  \mathcal{K}_t(X) )dW(t) . 
 \label{eq:pi-dyn-a-00}
\end{eqnarray}
Here,
\begin{equation}
  \mathcal{K}_t(X) =  \pi^{11}_t(L+L^\ast) + \pi^{01}_t(S) \xi(t) +
\pi^{10}_t(S^\ast) \xi^\ast(t)  
\end{equation}
and
 the innovations process $W(t)$ is a Wiener process with
respect to the single photon state and is defined by
\begin{equation}
dW(t) = dY(t) -  \mathcal{K}_t(X) dt .  \label{eq:innovation-11}
\end{equation}
We have $\pi^{01}_t(X) = \pi^{10}_t(X^\ast)^\ast$, and
the initial conditions are
$
\pi^{11}_0(X)= \pi^{00}_0(X)= \langle \eta, X \eta \rangle, \ \
\pi^{10}_0(X)= \pi^{01}_0(X)=0.
$

Consequently the conditional expectation for the system driven by the single photon field is given by
\begin{equation}
\pi_t(X) = \mathbb{E}_{\eta\xi} [ X(t) \vert Y(s), \  0 \leq s \leq t]  = \pi^{11}_t(X),
\label{eq:photon-pi11-def}
\end{equation}
and so the required quantum filter is given by the system of coupled equations (\ref{eq:pi-dyn-a-11})-(\ref{eq:pi-dyn-a-00}). 

Equations for the conditional density operators may easily be derived. Finally, we remark that although the master equations for systems driven by a single photon field have been obtained by other means in \cite{GEPZ98}, to the best of our knowledge the quantum filtering (trajectory) equations for such systems have not been derived before.

\section{Example}
\label{sec:photon-eg}

When the system  is a two-level system or qubit, the filtering equations
reduce to a finite set of stochastic differential equations. In this case we
have $\mathfrak{H}_S=\mathbb{C}^2$.
The system is specified by the parameters $S=I$, $L=\sqrt{\kappa}\, \sigma_-$, and $H=\omega \sigma_z$. Here $\kappa > 0$ is a scalar parameter.


We begin with the master equations (\ref{eq:rho-dyn-11})-(\ref{eq:rho-dyn-00}%
), and write
\begin{eqnarray}
\rho^{00} &=& \frac{1}{2} (I + x^{00} \sigma_x + y^{00} \sigma_y + z^{00}
\sigma_z )  \label{eq:rho-00} \\
\rho^{01} &=& \frac{1}{2} ( x^{01} \sigma_x + y^{01} \sigma_y + z^{01}
\sigma_z ) = \rho^{10\ast}  \label{eq:rho-01} \\
\rho^{11} &=& \frac{1}{2} (I + x^{11} \sigma_x + y^{11} \sigma_y + z^{11}
\sigma_z )  \label{eq:rho-11}
\end{eqnarray}
Note that $x^{00}$, $y^{00}$, $z^{00}$ and $x^{11}$, $y^{11}$, $z^{11}$ are
real, while $x^{01}$, $y^{01}$, $z^{01}$ may be complex. Also note, for
example, $\rho^{00}(\sigma_x) = x^{00}, \ \ \rho^{01}(\sigma_x) = x^{01\ast}$%
, etc. Then we obtain nine coupled equations for the nine coefficients:
\begin{align*}
\dot{x}^{00}&= -2\omega y^{00} -\frac{\kappa}{2}x^{00},\\
\dot{y}^{00} &= 2 \omega x^{00} - \frac{\kappa}{2}y^{00},\\
\dot{z}^{00} &=  -\kappa (1 +z^{11}),\\
\dot{x}^{01}&= -\frac{\kappa}{2}x^{01} - 2\omega y^{01} -\sqrt{\kappa}\xi(t)^* z^{00}, \\
\dot{y}^{01} &= 2 \omega x^{01} - \frac{\kappa}{2}y^{01}-i\sqrt{\kappa} \xi(t)^* z^{00},\\
\dot{z}^{01} &= -\kappa z^{01} -\sqrt{\kappa}x^{00} \xi(t)^* + i\sqrt{\kappa}y^{00}\xi(t)^*,  \\
\dot{x}^{11}&= -\frac{\kappa}{2}x^{11} - 2\omega y^{11} + \sqrt{\kappa}z^{01}\xi(t)
+\sqrt{\kappa}z^{01*}\xi(t)^*,\\
\dot{y}^{11} &= 2\omega x^{11} -\frac{\kappa}{2}y^{11} +i \sqrt{\kappa}z^{01}\xi(t)-i\sqrt{\kappa}z^{01*}\xi(t)^*, \\
\dot{z}^{11} &= -\kappa  -\kappa z^{11} -\sqrt{\kappa} x^{01} \xi(t) -i\sqrt{\kappa}y^{01}\xi(t)
\\
& \ \  - \sqrt{\kappa} x^{01*}\xi(t)^*+i\sqrt{\kappa}y^{01*}\xi(t)^*.
\end{align*}

For the quantum filter (\ref{eq:pi-dyn-a-11})-(\ref{eq:pi-dyn-a-00}),
we use a slightly more general representation for $\hat \rho^{jk}$ given by:
\begin{equation*}
\hat \rho^{jk} = \frac{1}{2}(\hat c^{jk}I + \hat x^{jk} \sigma_x + \hat y^{jk} \sigma_y+\hat z^{jk} \sigma_z),
\end{equation*} 
for $j,k=0,1$. Since $\langle \hat \rho^{11},I\rangle=1$ (i.e., $\hat \rho^{11}$ is a normalized conditional density operator), we always have that $\hat c^{11}=1$ at all times. However, unlike the master equation, this will not be so for $\hat c^{01},\hat c^{10},\hat c^{00}$, as these coefficients will evolve in time. This is the reason we need to consider the more general representation for $\hat \rho^{jk}$. The quantum filter for the two-level system is given by the finite set of coupled equations
\begin{align*}
d \hat c^{00} &=(\sqrt{\kappa}\hat x^{00} 
\\
& \ \ - (\sqrt{\kappa}\hat x^{11} +\frac{1}{2}\hat c^{01}\xi(t) + \frac{1}{2}\hat c^{01*}\xi(t)^*) \hat c^{00})dW(t),
\end{align*}
\begin{align*}
d\hat x^{00} &= \bigl(-2\omega y^{00} -\frac{\kappa}{2}x^{00}\bigr)dt 
 +\biggl(\sqrt{\kappa}\hat c^{00}
 \\
 & \ \ -(\sqrt{\kappa}\hat x^{11} +\frac{1}{2}\hat c^{01}\xi(t) + \frac{1}{2}\hat c^{01*}\xi(t)^*)\hat x^{00}\biggr)dW(t),
 \end{align*}
\begin{align*}
d\hat y^{00} &= \bigl(2 \omega x^{00} - \frac{\kappa}{2}y^{00}\bigr)dt 
\\
& \  \ - (\sqrt{\kappa}\hat x^{11} +\frac{1}{2}\hat c^{01}\xi(t) + \frac{1}{2}\hat c^{01*}\xi(t)^*)\hat y^{00} \biggr)dW(t),
\end{align*}
\begin{align*}
d\hat z^{00} &=  \bigl(-\kappa (1 +z^{00}))dt+ 
 \biggl(\sqrt{\kappa}\hat x^{00} 
 \\
 & \ \ - (\sqrt{\kappa}\hat x^{11} +\frac{1}{2}\hat c^{01}\xi(t) + \frac{1}{2}\hat c^{01*}\xi(t)^*) \hat z^{00}\biggr)dW(t),
\end{align*}
\begin{align*}
d \hat c^{01} & = \biggl(\sqrt{\kappa} \hat x^{01} + \hat c^{00} \xi(t)^* 
\\
& \  
- (\sqrt{\kappa}\hat x^{11} +\frac{1}{2}\hat c^{01}\xi(t) + \frac{1}{2}\hat c^{01*}\xi(t)^*)\hat c^{01}\biggr)dW(t),
\end{align*}
\begin{align*}
d\hat x^{01}&= \bigl(-\frac{\kappa}{2}x^{01} - 2\omega y^{01} -\sqrt{\kappa}\xi(t)^* z^{00}\bigr)dt 
   \\
 & \ \ +\biggl(\hat x^{00} \xi(t)^* 
 + \sqrt{\kappa} \hat c^{01}  
   \\
 & \ \ -(\sqrt{\kappa}\hat x^{11} 
 +\frac{1}{2}\hat c^{01}\xi(t) + \frac{1}{2}\hat c^{01*}\xi(t)^*) \hat x^{01}\biggr) dW(t),
 \end{align*}
\begin{align*}
d\hat y^{01} &= \bigl(2 \omega x^{01} - \frac{\kappa}{2}y^{01}-i\sqrt{\kappa} \xi(t)^* z^{00}\bigr)dt \\
&\quad +\biggl(\hat y^{00} \xi(t)^* 
   \\
 & \ \
 -(\sqrt{\kappa}\hat x^{11} +\frac{1}{2}\hat c^{01}\xi(t) + \frac{1}{2}\hat c^{01*}\xi(t)^*)\hat y^{01}\biggr) dW(t),
 \end{align*}
\begin{align*}
d\hat z^{01} &=  \bigl(-\kappa z^{01} -\sqrt{\kappa}x^{00} \xi(t)^* + i\sqrt{\kappa}y^{00}\xi(t)^* \bigr)dt \\
&\quad + \biggl(\sqrt{\kappa} \hat x^{01}
+ \hat z^{00} \xi(t)^*
   \\
 & \ \
 -(\sqrt{\kappa}\hat x^{11} +\frac{1}{2}\hat c^{01}\xi(t) + \frac{1}{2}\hat c^{01*}\xi(t)^*) \hat z^{01}\biggr)dW(t),
\end{align*}
\begin{align*}
d \hat x^{11}&= \bigl(-\frac{\kappa}{2}x^{11} - 2\omega y^{11} + \sqrt{\kappa}z^{01}\xi(t)
+\sqrt{\kappa}z^{01*}\xi(t)^*\bigr)dt \\
&\quad + \biggl(\sqrt{\kappa}+\hat x^{01*} \xi(t)^*+ \hat x^{01} \xi(t) 
   \\
 & \ \
 - (\sqrt{\kappa}\hat x^{11} +\frac{1}{2}c^{01}\xi(t) + \frac{1}{2}c^{01*}\xi(t)^*)\hat x^{11}\biggr)dW(t),
 \end{align*}
\begin{align*}
d \hat y^{11} &= \bigl(2\omega x^{11} -\frac{\kappa}{2}y^{11} +i \sqrt{\kappa}z^{01}\xi(t)-i\sqrt{\kappa}z^{01*}\xi(t)^* \bigr)dt \\
&\quad +\biggl(\hat y^{01*} \xi(t)^*+ \hat y^{01} \xi(t)
   \\
 & \ \
 -(\sqrt{\kappa}\hat x^{11} +\frac{1}{2}c^{01}\xi(t) + \frac{1}{2}c^{01*}\xi(t)^*)\hat y^{11}\biggr)dW(t),
 \end{align*}
\begin{align*}
d \hat z^{11} &= \bigl(-\kappa  -\kappa z^{11} 
 -\sqrt{\kappa} x^{01} \xi(t) -i\sqrt{\kappa}y^{01}\xi(t) 
    \\
 & \ \
 - \sqrt{\kappa} x^{01*}\xi(t)^*+i\sqrt{\kappa}y^{01*}\xi(t)^* \bigr)dt  
    \\
 & \ \
 +\biggl(\sqrt{\kappa}\hat x^{11} +\hat z^{01*} \xi(t)^*+ \hat z^{01} \xi(t)
   \\
 & \ \ -(\sqrt{\kappa}\hat x^{11} +\frac{1}{2}c^{01}\xi(t) + \frac{1}{2}c^{01*}\xi(t)^*)\hat z^{11}\biggr)dW(t).
\end{align*}
The innovations process is given by
\begin{equation}
dW(t) = dY(t)- \bigl(\sqrt{\kappa}\, \hat x^{11}(t) + \hat c^{01}(t)  \xi(t) + \hat c^{10}(t) \xi^\ast(t) \bigr)dt .
\end{equation}

\section{Discussion and Conclusion}
\label{sec:conclusion}

In this paper we have derived the master equation and quantum filter  for a class of open quantum systems
that are coupled to single photon   fields. The paper has focused on the case of quadrature measurements $Y(t)$ given by (\ref{eq:Y-out}), (\ref{eq:Z-def}). However, the methodology also applies to the case of counting measurements, corresponding to a photodetector in place of the homodyne detector in Figure \ref{fig:filter-one-1}.\footnote{Of course, homodyne detection is based on a photon counting system, e.g. \cite{GZ00}.}

The single photon filter   consist of   coupled equations  that determine the evolution of the conditional state of the system under continuous (weak) measurement performed on the output field, in contrast to 
the familiar single filtering equation for open Markov quantum systems that are coupled to coherent boson fields. 
 This coupled equations structure of the master and filter equations is a reflection of the non-Markov nature of systems coupled to single photon fields. 
  Indeed, a key feature of our approach is the  embedding of the system into a larger extended system, a technique often employed in the analysis of non-Markov  systems, providing an elegant framework within which to study the the dynamics, both unconditional and conditional, of the system.
 We expect that the basic approach taken in this paper can be adapted to study quantum systems that are coupled to other types of highly non-classical boson fields.

\appendix

The quantum filter for a system $G=(S,L,H)$ (driven by vacuum) for the quadrature output field observable $Y(t)$ (given by (\ref{eq:Y-out})) is  \begin{eqnarray}
& d\pi_t(X) =  \pi_t(   \mathcal{G}_G(X)) dt 
\label{eq:quant-filter-1}  \\
 & + ( \pi_t( XL + L^\ast X) - \pi_t(L+L^\ast) \pi_t(X)) d   W(t),
\notag
\end{eqnarray}
where $W(t)$, a Wiener process called the {\em innovations process}, is given by 
$
 dW(t) = dY(t) - \pi_t(L+L^\ast)dt).
$
By the spectral theorem it follows  that the quantum filter is equivalent to a classical system driven by the measurement record, \cite{VPB93,BHJ07}. Perhaps the simplest way to derive  the quantum filter   is to use the conditional characteristic function,  \cite{VPB92,VPB93,HSM05}, which we now briefly summarize. 

We begin with a short discussion of quantum conditional expectation. 
The measurement signal $Y(s)$, $0 \leq s \leq t$, is a collection of commuting self-adjoint operators. These operators form a subspace $\mathscr{Y}_t$ in the space of operators, and the quantum conditional expectation $\mathbb{E}_{\eta \phi}[ X(t) \vert Y(s)$, $0 \leq s \leq t]$ is the orthogonal projection of the system  operator $X(t)$ at time $t$ (since the field serves as a probe, we have the commutation relation  $[X(t), Y(s)]=0$ for all $0 \leq s \leq t$ (non-demolition), and so the conditional expectation is well-defined).  The orthogonal projection property corresponds to least squares estimation, and leads to the following characterization (\ref{eq:c-exp-def}) mentioned in Section \ref{sec:problem}.
We will use this characterization in the following form. 
Define, for any function $g$, 
\begin{equation}
c_g(t) = \exp\{   \int_0^t g(s) dY(s) - \frac{1}{2} \int_0^t \vert g(s) \vert^2 ds \},
\end{equation}
and note
$dc_g(t) = g(t) c_g(t) dY(t)$. Then we require
\begin{equation}
\mathbb{E}_{ \eta \phi } [ X(t) c_g(t) ] = \mathbb{E}_{\eta \phi} [ \hat X(t) c_g(t) ],
\label{eq:g}
\end{equation}
for all functions $g$.

Now suppose that $d\hat X(t)$ has the form
\begin{equation}
d\hat{X}(t)= \alpha(t) dt + \beta(t) dY(t),
\label{eq:hat-X-form}
\end{equation}
where $\alpha$ and $\beta$ are to be determined from the relation (\ref{eq:g}).
Now using the QSDE (\ref{eq:X-dyn}) for $X(t)=j_t(X)$,     we have
\begin{eqnarray}
&  \frac{d}{dt} \mathbb{E}_{\eta \phi} [ X(t) c_g(t) ]  
\notag
\\
&  =  
  \mathbb{E}_{\eta \phi} [ c_g(t) \pi_t(-i [X, H] + \mathcal{G}_L(X) ) 
\notag   \\
 &   + g(t) c_g(t) \pi_t( XL + L^\ast X) ] 
 \label{eq:filter-derive-1}
\end{eqnarray}
(here we have used property (\ref{eq:g})).
Similarly, using (\ref{eq:hat-X-form}) we have
\begin{eqnarray}
 & \frac{d}{dt} \mathbb{E}_{\eta \phi} [ \hat X(t) c_g(t) ]  
\notag \\
&=
 \mathbb{E}_{\eta \phi} [ c_g(t) (\alpha(t) + \pi_t(L+L^\ast) \beta(t))
 \notag \\
 & + g(t) c_g(t)( \beta(t) +\pi_t( L+L^\ast) \hat X(t) ]
 \label{eq:filter-derive-2}
\end{eqnarray}
Now equating the RHS of (\ref{eq:filter-derive-1}) and (\ref{eq:filter-derive-2}) and using the fact that $g$ is arbitrary we find that
 \begin{eqnarray*}
\alpha(t) &=& \pi_t( -i [X, H] + \mathcal{G}_L(X)) - \pi_t( L+L^\ast) \beta(t)
\\
\beta(t) &=& \pi_t( XL + L^\ast X) - \pi_t(L+L^\ast) \pi_t(X)
\end{eqnarray*}
The  quantum filter (\ref{eq:quant-filter-1}) now follows from this.

\section*{Acknowledgement}
The authors wish to thank J.~Hope for helpful discussions and for pointing out reference \cite{HPB04} to us. We also wish to thank A.~Doherty, H.~Wiseman, E.~Huntington  for helpful discussions and suggestions.

\bibliographystyle{plain}

\bibliography{mjbib2004}

\end{document}